\documentclass[a4paper,12pt]{article}
\usepackage[dvips]{graphicx}
\usepackage[centertags]{amsmath}
\usepackage{amsfonts}
\usepackage{amssymb}
\usepackage{latexsym}
\usepackage{mathrsfs}
\usepackage{cite}

\RequirePackage{ifpdf} %% We are running pdfTeX in pdf mode
\ifpdf
  \RequirePackage[%
    unicode,%
    pdfstartview=FitH,%
    colorlinks,%
    hyperfootnotes=true,%
    bookmarksopen,%
    bookmarksnumbered,%
    bookmarksopenlevel=3%
  ]{hyperref}
  \RequirePackage{thumbpdf}
\else
  \RequirePackage{hyperref}
\fi \RequirePackage{hyperref}

\newcommand{\email}[1]{\href{mailto:#1}{\texttt{#1}}}

\newcommand{\ket}[1]{\ensuremath{\left|#1\right>}}
\newcommand{\bra}[1]{\ensuremath{\left<#1\right|}}

\newcommand{\braket}[2]{\ensuremath{\left\langle{#1}\!%
\mathrel{\left|{\vphantom{{#1} {#2}}}\right.%
\kern-\nulldelimiterspace}\!{#2}\right\rangle}}

\newcommand{\BK}[3]{\ensuremath{\left\langle{#1}\!%
\mathrel{\left|\vphantom{{#1}}{#2}\vphantom{{#3}}\right|%
\kern-\nulldelimiterspace}\!{#3}\right\rangle}}

\newcommand{\commute}[2]{\ensuremath{\left[{#1}\!%
\mathrel{\vphantom{{#1}},\vphantom{{#2}}%
\kern-\nulldelimiterspace}\!{#2}\right]}}

\newcommand{\uI}{\ensuremath{\mathrm{i}}}
\newcommand{\uE}{\ensuremath{\mathrm{e}}}
\newcommand{\uD}{\ensuremath{\mathrm{d}}}

\renewcommand{\vec}[1]{\ensuremath{\boldsymbol{\mathrm{#1}}}}

\makeatletter
\let\@afterindentfalse\@afterindenttrue
\@afterindenttrue \makeatother
\begin{document}

%%\allowdisplaybreaks

%%\sloppy

\title{\textbf{Functional treatment of quantum scattering via the dynamical principle}}

\author{\textbf{Edouard~ B.~Manoukian}\footnote{Corresponding author.
e-mail:~\email{manoukian\_eb@hotmail.com}.} \ and \
\textbf{Seckson Sukkhasena} \\
{School of Physics, \ Institute of Science,} \\
{Suranaree University of Technology,} \\
{Nakhon~Ratchasima, 30000, Thailand}}

%\date{\today}

\maketitle

\begin{abstract}
A careful functional treatment of quantum scattering is given
using Schwinger's dynamical principle which involves a functional
differentiation operation applied to a generating functional
written in closed form. For long range interactions, such as for
the Coulomb one, it is shown that this expression may be used to
obtain explicitly the asymptotic ``free" modified Green function
near the energy shell. \\

\noindent\textbf{KEY~WORDS:} quantum dynamical principle, quantum scattering,
                            long range potentials, Green functions. 
\end{abstract}
\section{Introduction}\label{Section1}
The purpose of this communication is to use Schwinger's
\cite{Schwinger_1951,Schwinger_1953,Schwinger_1960,Schwinger_1962,Manoukian_1985}
most elegant quantum dynamical principle to provide a careful
functional treatment of quantum scattering. We derive rigorously
an expression for the scattering amplitude involving a functional
differentiation operation applied to a functional, depending on
the potential, written in closed form. The main result of this
paper is given in Eq.(\ref{Eqn2.28}). In particular, it provides a
systematic starting point for studies of deviations from so-called
straight-line ``trajectories'' of particles, with small deviation
angles, by mere functional differentiations. An investigation of a
time limit of a function related to this expression shows that the
latter may be also used to obtain the asymptotic ``free" modified
Green functions for theories with long range potentials such as
for the Coulomb potential with the latter defining the
transitional potential between short and long range potentials.
Functional methods have been also introduced earlier in the
literature
\cite{Brenner_1982,Chuluunbaatar_2001,Cambell_1975,Gelman_1969
,Gerry_1980,Pazma_1979,
Singh_1975,Zubarev_1977,Zubarev_1978,Sukumar_1984} in quantum
scattering dealing with path integrals or variational optimization
methods which, however, are not in the spirit of the present paper
based on the dynamical principle. The present study is an
adaptation of quantum field theory methods \cite{Manoukian_1988}
to quantum potential scattering.
%-------------------------
%\setcounter{page}{1}
\setcounter{section}{1}
%\renewcommand{\theequation}{\thechapter.\arabic{equation}}
%\numberwithin{equation}{chapter}
%-------------------------
%\renewcommand{\thesection}{\thechapter.\arabic{section}}
\renewcommand{\theequation}{\thesection.\arabic{equation}}
\numberwithin{equation}{section}
%--------------------------

\section{Functional treatment of scattering}\label{Section2}

Given a Hamiltonian
\begin{equation}\label{Eqn2.1}
H=\frac{\vec{p}^2}{2m}+V(\vec{x})
\end{equation}
for a particle of mass \textit{m} interacting with a potential
$V(\vec{x})$, we introduce a Hamiltonian $H'(\lambda,\tau)$
involving external sources $\vec{F}(\tau)$, $\vec{S}(\tau)$
coupled linearly to $\vec{x}$ and $\vec{p}$ as follows:
\begin{equation}\label{Eqn2.2}
H'(\lambda,\tau) = \frac{\vec{p}^2}{2m}+\lambda
V(\vec{x})-\vec{x}\cdot\vec{F}(\tau) + \vec{p}\cdot\vec{S}(\tau)
\end{equation}
where $\lambda$ is an arbitrary parameter which will be eventually
set equal to one. Schwinger's
~(\cite{Schwinger_1951}-\cite{Schwinger_1962},
\cite{Manoukian_1985}) dynamical principle states, that the
variation of the transformation function
$\braket{\vec{x}t}{\vec{p}t'}$ with respect to the parameter
$\lambda$ for the theory governed by the Hamiltonian
$H'(\lambda,\tau)$ is given by
\begin{equation}\label{Eqn2.3}
\delta\braket{\vec{x}t}{\vec{p}t'}=\left(-\dfrac{\uI}{\hbar}\right)
\int_{t'}^t \uD\tau\,\delta\left(\lambda
V\left(-\uI\hbar\dfrac{\delta}{\delta\vec{F}(\tau)}\right)\right)\braket{\vec{x}t}{\vec{p}t'}.
\end{equation}

Here $V(-\uI\hbar\delta/\delta\vec{F}(\tau))$ denotes $V(\vec{x})$
with $\vec{x}$ in it replaced by
$-\uI\hbar\delta/\delta\vec{F}(\tau)$. Eq.(\ref{Eqn2.3}) may be
readily integrated for ~$\lambda=1$,~$\vec{F}(\tau)$,
$\vec{S}(\tau)$ set equal to zero, that is for the theory governed
by the Hamiltonian $H$ in (\ref{Eqn2.1}), to obtain

\begin{equation}\label{Eqn2.4}
\braket{\vec{x}t}{\vec{p}t'} =
\exp\left[-\dfrac{\uI}{\hbar}\int_{t'}^t \uD\tau{}\,
V\left(-\uI\hbar\dfrac{\delta}{\delta\vec{F}(\tau)}\right)\right]
\braket{\vec{x}t}{\vec{p}t'}^{(0)}\bigg|_{\vec{F}=0,\vec{S}=0}
\end{equation}

The transformation function $\braket{\vec{x}t}{\vec{p}t'}^{(0)}$
corresponds a theory developing in time via the Hamiltonian

\begin{equation}\label{Eqn2.5}
H'(0,\tau) = \dfrac{\vec{p}^2}{2m}-\vec{x}\cdot\vec{F}(\tau) +
\vec{p}\cdot\vec{S}(\tau),
\end{equation}
to which we now pay special attention.

With $\vec{p}$ replaced by $\uI\hbar\delta/\delta\vec{S}(\tau)$,
the dynamical principle, exactly as in (\ref{Eqn2.4}), gives
\begin{equation}\label{Eqn2.6}
\braket{\vec{x}t}{\vec{p}t'}^{(0)}=\exp\left[-\dfrac{\uI}{2m\hbar}\int_{t'}^t
\uD\tau
\left(\uI\hbar\dfrac{\delta}{\delta\vec{S}(\tau)}\right)^2\right]\braket{\vec{x}t}{\vec{p}t'}_0
\end{equation}
where the transformation function $\braket{\vec{x}t}{\vec{p}t'}_0$
is governed by the ``Hamiltonian"
\begin{equation}\label{Eqn2.7}
\hat{H}(\tau) = -\vec{x}\cdot\vec{F}(\tau) +
\vec{p}\cdot\vec{S}(\tau),
\end{equation}
involving no kinetic energy term.

The Heisenberg equations corresponding to $\Hat{H}(\tau)$ give the
equations
\begin{equation}\label{Eqn2.8}
\vec{x}(\tau)= \vec{x}(t) - \int_{t'}^t\uD\tau'\,
\Theta(\tau'-\tau)\vec{S}(\tau'),
\end{equation}
\begin{equation}\label{Eqn2.9}
\vec{p}(\tau)= \vec{p}(t') + \int_{t'}^t\uD\tau'\,
\Theta(\tau-\tau')\vec{F}(\tau'),
\end{equation}
where $\Theta (\tau)$ is the step function $\Theta (\tau)= 1$ for
$\tau > 0$ and $=0$ for $\tau < 0$. Using the relations
\begin{equation}\label{Eqn2.10}
_0\negmedspace\bra{\vec{x}t}\vec{x}(t) = \vec{x}\bra{\vec{x}t},
\end{equation}
\begin{equation}\label{Eqn2.11}
\vec{p}(t')\ket{\vec{p}t'}_0 = \ket{\vec{p}t'}\vec{p},
\end{equation}
and the dynamical principle, we obtain from taking the matrix
elements of $\vec{x}(\tau)$,~$\vec{p}(\tau)$ in
(\ref{Eqn2.8}),(\ref{Eqn2.9}) between the states
$_0\negmedspace\bra{\vec{x}t}$~,~$\ket{\vec{p}t'}_0$,~the
functional differential equations
\begin{equation}\label{Eqn2.12}
-\uI\hbar\dfrac{\delta}{\delta\vec{F}(\tau)}\braket{\vec{x}t}{\vec{p}t'}_0
= \left[\vec{x} - \int_{t'}^t\uD\tau'\,
\Theta(\tau'-\tau)\vec{S}(\tau')\right]\braket{\vec{x}t}{\vec{p}t'}_0
\end{equation}
\begin{equation}\label{Eqn2.13}
\uI\hbar\dfrac{\delta}{\delta\vec{S}(\tau)}\braket{\vec{x}t}{\vec{p}t'}_0
=\left[\vec{p} + \int_{t'}^t\uD\tau'\,
\Theta(\tau-\tau')\vec{F}(\tau')\right]\braket{\vec{x}t}{\vec{p}t'}_0
\end{equation}

These equations may be integrated to yield
\begin{eqnarray}\label{Eqn2.14}
\braket{\vec{x}t}{\vec{p}t'}_0 =
\exp\left[\dfrac{\uI}{\hbar}\vec{x}\cdot\left(\vec{p}+ \int_{t'}^t
\uD\tau \,\vec{F}(\tau)\right)\right]
\exp\left[-\dfrac{\uI}{\hbar}\vec{p}\cdot\int_{t'}^t\uD\tau\,
\vec{S}(\tau)\right] \nonumber\\ \times
\exp\left[-\dfrac{\uI}{\hbar}\int_{t'}^t\uD\tau
\int_{t'}^t\uD\tau'\,\vec{S}(\tau)\cdot\vec{F}(\tau')\Theta(\tau-\tau')
\right],
\end{eqnarray}
satisfying the familiar boundary condition $\exp
(\uI\vec{x}\cdot\vec{p}/\hbar)$ for $\vec{F},\, $\vec{S} set equal
to zero.

Since we are interested in (\ref{Eqn2.4}), in particular, for the
case when $\vec{S}$ is set equal to zero, the functional
differentiation in (\ref{Eqn2.6}) may easily carried out giving
\begin{align}\label{Eqn2.15}
\braket{\vec{x}t}{\vec{p}t'}^{(0)}\bigg|_{\vec{S}=\vec{0}} &=
\exp\left[\dfrac{\uI}{\hbar}\left(\vec{x}\cdot\vec{p}-\dfrac{\vec{p}^2}{2m}(t-t')\right)\right]\nonumber\\
&\times\exp\left[\dfrac{\uI}{\hbar}\int_{t'}^t\uD\tau\,
\vec{F}(\tau)\cdot\left(\vec{x}-\dfrac{\vec{p}}{m}(t-\tau)\right)\right]\nonumber\\
&\times\exp\left[-\dfrac{\uI}{2m\hbar}\int_{t'}^t\uD\tau\int_{t'}^t\uD\tau'\,
\vec{F}(\tau)\cdot\vec{F}(\tau')(t-\tau_>)\right],
\end{align}
where $\tau_>$ is the largest of $\tau$ and $\tau' :
\tau=max(\tau,\tau')$.

We recall from (\ref{Eqn2.4}) that we eventually set
$\vec{F}(\tau)$ equal to zero. This allows us to interchange the
exponential factor in (\ref{Eqn2.4}) involving the
$V(-\uI\hbar\delta/\delta\vec{F}(\tau))$ term and the last two
exponential factors in (\ref{Eqn2.15}). This gives for
$\braket{\vec{x}t}{\vec{p}t'}$ in (\ref{Eqn2.4}) the expression
\begin{align}\label{Eqn2.16}
\braket{\vec{x}t}{\vec{p}t'}&= \exp\left[\dfrac{\uI}{\hbar}
\left(\vec{x}\cdot\vec{p}-\dfrac{\vec{p}^2}{2m}(t-t')\right)\right]\nonumber\\
&\times
\exp\left[\dfrac{\uI\hbar}{2m}\int_{t'}^t\uD\tau\int_{t'}^t\uD\tau'\left[t-\tau_>\right]
\dfrac{\delta}{\delta\vec{F}(\tau)}
\cdot\dfrac{\delta}{\delta\vec{F}(\tau')}\right]\nonumber\\
&\times \exp\left[-\dfrac{\uI}{\hbar}\int_{t'}^t\uD\tau\,
V\left(\vec{x}-\dfrac{\vec{p}}{m}(t-\tau)+\vec{F}(\tau)\right)\right]\bigg|_{\vec{F}=\vec{0}}.
\end{align}

Since we finally set $\vec{F}= \vec{0}$ in (\ref{Eqn2.16}), the
theory becomes translational invariant in time and
$\braket{\vec{x}t}{\vec{p}t'}$ is a function of $t-t'\equiv T$

For $t>t'$, we have the definition of the Green function
\begin{equation}\label{Eqn2.17}
\braket{\vec{x}t}{\vec{x}'t'}=G_+(\vec{x}t,\vec{x}'t'),
\end{equation}
with $G_+(\vec{x}t,\vec{x}'t')$=0 for $t<t'$, and
\begin{eqnarray}\label{Eqn2.18}
\braket{\vec{x}t}{\vec{p}t'}&=&G_+(\vec{x}t,\vec{p}t')\nonumber\\
&=&\int\uD^3\vec{x}'\,
\uE^{\uI\vec{p}\cdot\vec{x}'/\hbar}G_+(\vec{x}t,\vec{x}'t').
\end{eqnarray}

We may now introduce the Fourier transform defined by
\begin{equation}\label{Eqn2.19}
G_+(\vec{p},\vec{p}';p^0)=-\dfrac{\uI}{\hbar}\dfrac{1}{(2\pi\hbar)^3}\int_0^\infty\uD
T\, \uE^{\uI(p^0+\uI\epsilon)T/\hbar}\int\uD^3\vec{x}\,
\uE^{-\uI\vec{p}\cdot\vec{x}}\braket{\vec{x}T}{\vec{p}'0},
\end{equation}
for $\epsilon\rightarrow +0$, where $\braket{\vec{x}T}{\vec{p}0}$
is given in (\ref{Eqn2.16}) with $t-t'\equiv T$. From
(\ref{Eqn2.19}),~(\ref{Eqn2.16}), we may rewrite
$G_+(\vec{p},\vec{p}';p^0)$ as
\begin{equation}\label{Eqn2.20}
G_+(\vec{p},\vec{p}';p^0)=-\dfrac{\uI}{\hbar}\dfrac{1}{(2\pi\hbar)^3}\int_0^\infty\uD\alpha\,
\uE^{\uI[p^0 - E(\vec{p}')+
\uI\epsilon]\alpha/\hbar}\int\uD^3\vec{x}\, \uE^{-\uI\vec{x}\cdot
(\vec{p}-\vec{p}')/\hbar}K(\vec{x},\vec{p}';\alpha),
\end{equation}
where $E(\vec{p})=\vec{p}^2/2m$,
\begin{eqnarray}\label{Eqn2.21}
K(\vec{x},\vec{p}';\alpha)=
\exp\left[\dfrac{\uI\hbar}{2m}\int_{t'}^t\uD\tau
\int_{t'}^t\uD\tau'
[t-\tau_>]\dfrac{\delta}{\delta\vec{F}(\tau)}\cdot\dfrac{\delta}{\delta\vec{F}(\tau')}\right]
\nonumber\\
\times\exp\left[-\dfrac{\uI}{\hbar}\int_{t'}^t\uD\tau\,
V\left(\vec{x} - \dfrac{\vec{p}'}{m}(t-\tau) +
\vec{F}(\tau)\right)\right]\bigg|_{\vec{F}=\vec{0}}
\end{eqnarray}
\textit{with} $t-t'\equiv\alpha$ playing the role of time $-$ a
notation used for it quite often in field theory.

In the $\alpha$-integrand in the exponential in (\ref{Eqn2.20}),
we recognize $[p^0-E(\vec{p})  + \uI\epsilon]$ as the inverse of
the free Green function in the energy-momentum representation.

The scattering amplitude $f(\vec{p},\vec{p}')$ for scattering of
the particle with initial and final momenta $\vec{p}'$,~$\vec{p}$,
respectively, is defined by
\begin{equation}\label{Eqn2.22}
f(\vec{p},\vec{p}')=
-\dfrac{m}{2\pi\hbar^2}\int\uD^3\vec{p}''\,V(\vec{p}-\vec{p''})G_+(\vec{p}'',\vec{p}';p^0)[p^0-E(\vec{p}')]\bigg|_{p^0=E(\vec{p}')}
\end{equation}
where $V(\vec{p}) = \int \uD^3\vec{x}\,
\uE^{-\uI\vec{x}\cdot\vec{p}/\hbar}V(\vec{x})$. This suggests to
multiply (\ref{Eqn2.20}) by $[p^0-E(\vec{p}')]$ giving
\begin{align}\label{Eqn2.23}
G_+(\vec{p},\vec{p}';p^0)[p^0-E(\vec{p}')]&=-\dfrac{1}{(2\pi\hbar)^3}\int_0^\infty\uD\alpha
\left(\dfrac{\partial}{\partial\alpha}\uE^{\uI\alpha[p^0-E(\vec{p}')+\uI\epsilon]/\hbar}\right)\nonumber\\[0.5\baselineskip]
&\times\int\uD^3\vec{x}\,\uE^{-\uI\vec{x}\cdot
(\vec{p}-\vec{p}')/\hbar}K(\vec{x},\vec{p}';\alpha).
\end{align}

From the fact that $\braket{\vec{x}}{\vec{p}}=\exp
(\uI\vec{x}\cdot\vec{p}/\hbar)$ and the definition of
$K(\vec{x},\vec{p}';\alpha)$ in (\ref{Eqn2.21}), we have
\begin{equation}\label{Eqn2.24}
K(\vec{x},\vec{p}';0)=1.
\end{equation}

We now consider the cases for which
\begin{equation}\label{Eqn2.25}
\lim_{\alpha
\rightarrow\infty}\int\uD^3\vec{x}\,\uE^{-\uI\vec{x}\cdot(\vec{p}-\vec{p}')/\hbar}K(\vec{x},\vec{p}';\alpha),
\end{equation}
\textit{exists}. This, in particular, implies that $(\epsilon >
0)$
\begin{equation}\label{Eqn2.26}
\lim_{\alpha
\rightarrow\infty}\uE^{-\epsilon\alpha}\int\uD^3\vec{x}\,
\uE^{-\uI\vec{x}\cdot(\vec{p}-\vec{p}')/\hbar}
K(\vec{x},\vec{p}';\alpha)=0.
\end{equation}

We may then integrate over $\alpha$ in (\ref{Eqn2.23}) to obtain
simply
\begin{eqnarray}\label{Eqn2.27}
G_+(\vec{p},\vec{p}';p^0)[p^0-E(\vec{p}')]\bigg|_{p^0=E(\vec{p}')}\qquad\qquad\qquad\nonumber\\
= \lim_{\alpha
\rightarrow\infty}\dfrac{1}{(2\pi\hbar)^3}\int\uD^3\vec{x}\,
\uE^{-\uI\vec{x}\cdot (\vec{p}-\vec{p}')/\hbar}
K(\vec{x},\vec{p}';\alpha),
\end{eqnarray}
on the energy shell $p^0=E(\vec{p})$, and for the scattering
amplitude, in (\ref{Eqn2.22}), \textit{after} integrating over
$\vec{p}''$, the expression
\begin{equation}\label{Eqn2.28}
f(\vec{p},\vec{p}')=-\dfrac{m}{2\pi\hbar^2}\lim_{\alpha\rightarrow\infty}\int\uD^3\vec{x}\,
\uE^{-\uI\vec{x}\cdot
(\vec{p}-\vec{p}')/\hbar}V(\vec{x})K(\vec{x},\vec{p}';\alpha),
\end{equation}
with $K(\vec{x},\vec{p}';\alpha)$ defined in (\ref{Eqn2.21}). Here
we recognize that the formal replacement of
$K(\vec{x},\vec{p}';\alpha)$ by one gives the celebrated Born
approximation. On the other hand, part of the argument $[\vec{x} -
\vec{p}'(t-\tau)/m]$ ~of~ $V(\vec{x} - \vec{p}'(t-\tau)/m +
\vec{F}(\tau))$ in (\ref{Eqn2.21}), represents a ``straight line
trajectory" of a particle, with the functional differentiations
with respect to $\vec{F}(\tau)$, as defined in (\ref{Eqn2.21}),
leading to deviations of the dynamics from such a straight line
trajectory. With a straight line approximation, ignoring all of
the functional differentiations, with respect to $\vec{F}(\tau)$
and setting the latter equal to zero, gives the following explicit
expression for the scattering amplitude $f(\vec{p},\vec{p}')$ in
(\ref{Eqn2.28}):

\begin{equation}\label{Eqn2.29}
f(\vec{p},\vec{p}')=-\dfrac{m}{2\pi\hbar^2}\int\uD^3\vec{x}\,
\uE^{-\uI\vec{x}\cdot
(\vec{p}-\vec{p}')/\hbar}V(\vec{x})\exp\left[-\dfrac{\uI}{\hbar}\int_0^\infty
\uD\alpha\,
V\left(\vec{x}-\dfrac{\vec{p}'}{m}\alpha\right)\right].
\end{equation}

This modifies the Born approximation by the presence of an
additional phase factor in the integrand in (\ref{Eqn2.29}),
depending on the potential, accumulated during the scattering
process. Here one recognizes the expression which leads to
scattering with small deflections at high energies (the so-called
eikonal approximation) obtained from the straight line trajectory
approximation discussed above. Deviations from this approximation
may be then systematically obtained by carrying out a functional
power series expansion of $V\left(\vec{x} - \vec{p}'(t-\tau)/m +
\vec{F}(\tau)\right)$ in $\vec{F}(\tau)$ and performing the
functional differential operation as dictated by the first
exponential in (\ref{Eqn2.21}) and finally setting $\vec{F}(\tau)$
equal to zero.

We note that formally that the $\tau$-integral, involving the
potential $V$, in (\ref{Eqn2.21}) increases with no bound for
$\alpha\rightarrow\infty$ for the Coulomb potential and for
potentials of longer range with the former potential defining the
transitional potential between long and short range potentials.
And in case that the limit in (\ref{Eqn2.25}) does not exist, as
encountered for the Coulomb potential, (\ref{Eqn2.23}) cannot be
integrated by parts. This is discussed in the next section.

\section{Asymptotic ``free" Green function}\label{Section3}

In case the $\alpha \rightarrow \infty$ limit in (\ref{Eqn2.25})
does not exist, one may study the behaviour of
$G_+(\vec{p},\vec{p}';p^0)$ near the energy shell $p^0\simeq
\vec{p}'^2/2m$ directly from (\ref{Eqn2.20}). To this end, we
introduce the integration variable
\begin{equation}\label{Eqn3.1}
z=\dfrac{\alpha}{\hbar}\left[p^0 - E(\vec{p}')\right],
\end{equation}
in (\ref{Eqn2.20}), to obtain
\begin{align}\label{Eqn3.2}
G_+(\vec{p},\vec{p}';p^0)[p^0-E(\vec{p}')]=-\dfrac{\uI}{(2\pi\hbar)^3}
\int_0^\infty \uD  z\, \uE^{\uI z(1+\uI\epsilon)}\nonumber\\
\times \int\uD^3 \vec{x}\, \uE^{-\uI\vec{x}\cdot
(\vec{p}-\vec{p}')/\hbar}\,K\left(\vec{x},\vec{p}';\dfrac{z\hbar}{p^0-E(\vec{p}')}\right),
\end{align}
for $\epsilon\rightarrow +0$. For $p^0-E(\vec{p}')\gtrsim 0$,
i.e., near the energy shell, we may substitute
\begin{equation}\label{Eqn3.3}
K\left(\vec{x},\vec{p}';z\hbar/(p^0-E(\vec{p}'))\right)\simeq
\exp\left[-\dfrac{\uI}{\hbar}
\int_0^{z\hbar/(p^0-E(\vec{p}'))}\uD\alpha\,
V\left(\vec{x}-\dfrac{\vec{p}'}{m}\alpha\right)\right],
\end{equation}
in (\ref{Eqn3.2}) to obtain for the following integral
\begin{eqnarray}\label{Eqn3.4}
\int\uD^3\vec{p}\,
\uE^{\uI\vec{p}\cdot\vec{x}/\hbar}G_+(\vec{p},\vec{p}';p^0)\simeq
    \dfrac{-\uI\uE^{\uI\vec{x}\cdot\vec{p}'/\hbar}}{\left[p^0-E(\vec{p}')+\uI\epsilon\right]}\int_0^\infty\uD
    z\, \uE^{\uI z(1+\uI\epsilon)}\nonumber\\
    \times\exp\left[-\dfrac{\uI}{\hbar}\int_0^{z\hbar/(p^0-E(\vec{p}'))}\uD\alpha\,
    V\left(\vec{x}-\dfrac{\vec{p}'}{m}\alpha\right)\right],
\end{eqnarray}

For the Coulomb potential $V(\vec{x}) = \lambda/|\vec{x}|$,
\begin{equation}\label{Eqn3.5}
\int_0^{z\hbar/(p^0-E(\vec{p}'))}\uD\alpha\,
    V\left(\vec{x}-\dfrac{\vec{p}'}{m}\alpha\right)\simeq
    \dfrac{\lambda m}{|\vec{p}'|}\ln\left(\dfrac{2|\vec{p}'|z\hbar}{m(p^0-E(\vec{p}'))|\vec{x}|(1-\cos\theta)}\right)
\end{equation}
where $\cos\theta = \vec{p}'\cdot\vec{x}/|\vec{p}'||\vec{x}|$.
Hence
\begin{eqnarray}\label{Eqn3.6}
\exp
-\dfrac{\uI}{\hbar}\int_0^{z\hbar/(p^0-E(\vec{p}'))}\uD\alpha\,
    V\left(\vec{x}-\dfrac{\vec{p}'}{m}\alpha\right)\simeq
    \dfrac{1}{\left[p^0-E(\vec{p}')+\uI\epsilon\right]^{-\uI\gamma}}\nonumber\\
    \times\exp -\uI\gamma \ln\left(\dfrac{2p'^2{} z\hbar}{m(p'x -
    \vec{p}'\cdot\vec{x})}\right),
\end{eqnarray}
where $\gamma = \lambda m/\hbar p'$. Finally using the integral
\begin{equation}\label{Eqn3.7}
\int_0^\infty \uD z\, \uE^{\uI
z(1+\uI\epsilon)}(z)^{-\uI\gamma}=\uI
\uE^{\pi\gamma/2}\Gamma(1-\uI\gamma),
\end{equation}
for $\epsilon\rightarrow +0$,~where $\Gamma$ is the gamma
function, we obtain from (\ref{Eqn3.4})
\begin{eqnarray}\label{Eqn3.8}
\int\uD^3\vec{p}\,
\uE^{\uI\vec{p}\cdot\vec{x}/\hbar}G_+(\vec{p},\vec{p}';p^0)\simeq
\uE^{\uI\vec{x}\cdot\vec{p}'/\hbar}\dfrac{\uE^{-\uI\gamma\ln(2p'^2/m)}}{[p^0-E(\vec{p}')+\uI\epsilon]^{1-\uI\gamma}}\nonumber\\
    \times\exp\uI\gamma\ln\left(\dfrac{p'x-\vec{p}'\cdot\vec{x}}{\hbar}\right)
    \uE^{\pi\gamma/2}\Gamma(1-\uI\gamma),
\end{eqnarray}
to be compared with earlier results (e.g.,
\cite{Papanicolaou_1976}), and for the asymptotic ``free" Green
function, in the energy-momentum representation, the expression.
\begin{equation}\label{Eqn3.9}
G_+^0(\vec{p})=
\dfrac{\uE^{-\uI\gamma\ln(2p^2/m)}}{[p^0-E(\vec{p})+\uI\epsilon]^{1-\uI\gamma}}\uE^{\pi\gamma/2}\Gamma(1-\uI\gamma).
\end{equation}
showing on obvious modification from the Fourier transform of the
free Green function $[p^0-E(\vec{p})+\uI\epsilon]^{-1}$.

\section{Conclusion}\label{Section4}

The expression (\ref{Eqn2.28}) provides a functional expression
for the scattering amplitude with $K(\vec{x},\vec{p}';\alpha)$
defined in (\ref{Eqn2.21}) and the latter is obtained by the
functional differential operation carried out on the functional,
involving the potential $V$, of argument $\vec{x}-
\vec{p}'(t-\tau)/m + \vec{F}(\tau)$, for all $t'\leq\tau\leq t$,
represented by the first exponential in (\ref{Eqn2.21}). The
``straight line trajectory" approximation of a particle consisting
of retaining $\vec{x} - \vec{p}'(t-\tau)/m$ only in the argument
of $V$ and neglecting the functional differentiations with respect
to $\vec{F}(\tau)$, with the latter operation leading
systematically to modifications from this linear ``trajectory",
gives rise to the familiar eikonal approximation. The existence of
the time limit in (\ref{Eqn2.25}) distinguishes between so-called
potentials of short and long ranges with the Coulomb potential
providing the transitional potential between these two general
classes of potentials and belonging to the latter class. In case
the time limit in (\ref{Eqn2.25}) does not exist, corresponding to
potentials of long ranges, (\ref{Eqn2.20}) may be used to obtain
the asymptotic ``free" modified Green function near the energy
shell as seen in Sect.\ref{Section3}. In a subsequent report, our
functional expression in (\ref{Eqn2.28}) for the scattering
amplitude will be generalized for long range potentials as well.


\newpage


\begin{thebibliography}{99}
\raggedright

\bibitem{Brenner_1982}
Brenner,~S.~E., Galimzyanov,~R.~M. (1982). \textit{Czechoslovak
Journal of Physics B} ~\textbf{32}, 316.

\bibitem{Chuluunbaatar_2001}
Chuluunbaatar,~O., Puzynin,~I.~V. and Vinitoky,~S.~I. (2001).
\textit{Joint Institute for Nuclear Research} ~\textbf{P11},
2001-61.

\bibitem{Cambell_1975}
Cambell,~W.~B. (1975). \textit{Physical Review D} ~\textbf{12},
2362.

\bibitem{Gelman_1969}
Gelman,~D., Spruch,~L. (1969). \textit{Journal of Mathematical
Physics} ~\textbf{10}, 2240.

\bibitem{Gerry_1980}
Gerry,~C.~C. (1980). \textit{Physical Review D} ~\textbf{21},
2979.

\bibitem{Manoukian_1985}
Manoukian,~E.~B. (1985). \textit{Nuovo Cimento A}~\textbf{90},
295.
%%``Quantum Action Principle and Path Integrals for
%%Long-Range Interactions'',

\bibitem{Manoukian_1988}
Manoukian,~E.~B. (1988). \textit{International Journal of
Theoretical Physics} ~\textbf{27}, 401.

\bibitem{Papanicolaou_1976}
Papanicolaou,~N. (1976). \textit{Physics Reports}~\textbf{24},
229.
%%\href{http://dx.doi.org/10.1016/0370-2693(67)90067-6}
%%{``Feynman Diagrams for the Yang--Mills Field''},

\bibitem{Pazma_1979}
Pazma,~V. (1979). \textit{Acta Physica Slovaca} ~\textbf{29}, 249.

\bibitem{Schwinger_1951}
Schwinger,~J. (1951). \textit{Proceedings of the National Academy
of Sciences USA.}~\textbf{37}, 452.

\bibitem{Schwinger_1953}
Schwinger,~J. (1953). \textit{Physical Review} ~\textbf{91}, 713.

\bibitem{Schwinger_1960}
Schwinger,~J. (1960a,b). \textit{Proceedings of the National
Academy of Sciences USA.} ~\textbf{46}, 883,~\textbf{46}, 1401.

\bibitem{Schwinger_1962}
Schwinger,~J. (1962). \textit{Proceedings of the National Academy
of Sciences USA.} ~\textbf{48}, 663.

\bibitem{Singh_1975}
Singh,~S.~R. (1975). \textit{Journal of Physics A: Mathematical
and General} ~\textbf{8}, 1379.

\bibitem{Sukumar_1984}
Sukumar,~C.~V. (1984). \textit{Journal of Physics G: Nuclear
Physics} ~\textbf{10}, 81.
%------
\bibitem{Zubarev_1977}
Zubarev,~A.~L. (1977). \textit{Theoretical and Mathematical
Physics} ~\textbf{30}, 45.

\bibitem{Zubarev_1978}
Zubarev,~A.~L. (1978). \textit{Yadernaya fizika} ~\textbf{18},
566.
%------

%17

\end{thebibliography}
\end{document}